\documentclass[graybox]{svmult}
\usepackage{mathptmx}       
\usepackage{helvet}         
\usepackage{courier}        
\usepackage{type1cm}        
\usepackage{makeidx}         
\usepackage{graphicx}        
\usepackage{multicol}        
\usepackage[bottom]{footmisc}

\makeindex             

\begin{document}

\title*{The Dynamics of an Expanding OB Association}
\titlerunning{Dynamics of an OB Association}

\author{Nicholas J. Wright, Herve Bouy, Jeremy J. Drake, Janet E. Drew, Mario Guarcello, David Barrado y Navacu\'es}
\authorrunning{Nicholas J. Wright et al.}

\institute{Nicholas J. Wright \at Harvard-Smithsonian Center for Astrophysics \email{nwright@cfa.harvard.edu}
}

\maketitle

\vspace{-3cm}

\abstract{We present 3-dimensional kinematical observations of the massive OB association Cygnus~OB2 to identify the mechanisms responsible for disrupting young star clusters. The picture revealed by these observations is of a highly-substructured, dynamically unmixed OB association that does not exhibit the position--velocity correlations predicted by the theories of infant mortality or tidal stripping. These observations would appear to support a picture of hierarchical star formation.}

\section{Introduction}

It has been known for many years that there is a lack of old clusters compared to an extrapolation of the young cluster population. This was first recognized by Oort [17] and exacerbated by the discovery of large numbers of embedded clusters in the near-IR [13]. Based on this it became clear that the vast majority (at least 90\%) of clusters disperse within 10~Myr [14].

This is usually explained by the process of {\it infant mortality}, whereby residual gas left over from star formation is forced out of the cluster via feedback from massive stars, leaving the stellar part of the system in a super-virial state and prone to dissolution. This long-established theoretical framework assumes that the cluster was in virial equilibrium prior to gas expulsion and identifies the star formation efficiency (the fraction of gas turned into stars, $\sim$5-10\%) as the dominant factor in determining cluster stability [10,1]. This has recently been called into question: numerous theoretical works have suggested other parameters of equal importance such as the spatial distribution of stars at birth [15], the rate of residual gas expulsion [3], and the initial virial state of the cluster [16]. Overall, these studies suggest that the influence of gas expulsion may have been over-estimated and that many clusters may be stable from a young age.

An alternative theory, first put forward by Spitzer [18] to explain the lack of Gyr-old clusters, is that clusters are tidally heated by passing interstellar clouds. This has recently been used to explain the disruption of very young clusters by their parental giant molecular cloud (GMC) [7,12]. It has been argued to be an effective disruption mechanism for young clusters since the average cloud density in GMCs exceeds the tidal density from the galactic potential by an order of magnitude or more [7].

A third and important explanation for the lack of mature, gravitationally-bound clusters is that the majority of young clusters are not gravitationally bound, one of the fundamental assumptions of the theory of infant mortality. The discovery of vast numbers of near-IR embedded clusters led many authors to conclude that all stars formed in bound clusters [13,6,14]. However many of these clusters may just be stellar overdensities and may not be gravitationally bound. In their study of the spatial distribution of young stars, Bressert et al. [5] could find no evidence for a preferred scale for clustering that would be apparent if stars preferentially form in clusters (see also [8]). These findings support the view that star formation is hierarchical with no preference for scale and therefore most stars may not form in bound groups [2]. If the majority of stars do not form in bound clusters, then an efficient disruption mechanism may not be required.

To answer the question of how star clusters are disrupted, and thereby also address the question of whether all stars form in clusters, we must study a cluster in the process of being disrupted. This is difficult because the majority of clusters are either still embedded in their GMC (e.g. the Orion Nebular Cluster) or if they have already removed their residual gas they are often found to be gravitationally bound. This intrinsic bias is because clusters that display a clear overdensity and a spherical shape but which have also emerged from their parental GMC {\it must} be bound if they have retained their clustered morphology. To study a cluster in the act of dispersal and therefore probe the mechanisms responsible we should not study star clusters but instead study OB associations, less dense groups of stars that have been suggested to be the result of expanded clusters [13].

\section{Observations}

By studying the dynamics of an expanding OB association we intend to test theories for how star clusters are disrupted and probe the physical mechanisms at work. Radial velocities (RVs) are useful for this, but more important are proper motions (PMs) that provide a vital correlation between position and velocity that is necessary to distinguish between theories. For infant mortality we should see a radial dispersion of stars moving away from the original cluster center(s), whereas tidal heating predicts velocities distributed along a specific axis [12]. Hierarchical star formation should result in almost random motions based on the original spatial distribution.

We have targeted the massive OB association Cygnus~OB2, a post-gas expulsion association believed to be in the process of dispersing. It is the most massive group of young stars within 2~kpc, with a mass of $M_\star \sim 3 \times 10^4$~M$_\odot$ [21], similar to the most massive clusters in our galaxy. With an age of 3--5~Myr [9,21] it is old enough to be dynamically evolved, but with evidence for current star formation on the periphery [19,22]. Members of the association are selected using X-ray observations [20] since young stars are more X-ray luminous that main-sequence stars.

RVs were obtained from multi-epoch (to remove close binaries) MMT/Hectospec spectroscopy of the Ca~{\sc ii} triplet with an accuracy of $\sim$3--5~km/s. PMs are calculated from multi-epoch images spanning a 7--8~yr baseline using the DANCE (Dynamical Analysis of Nearby ClustErs, [4]) program. This results in an accuracy of $< 1$~mas/yr, equivalent to $\sim$3~km/s at the distance of Cyg~OB2. These velocities are sufficient to resolve internal substructure in a cluster with $\sigma_r \sim 10$~km/s [11].

\begin{figure}[t]
\includegraphics[scale=.645]{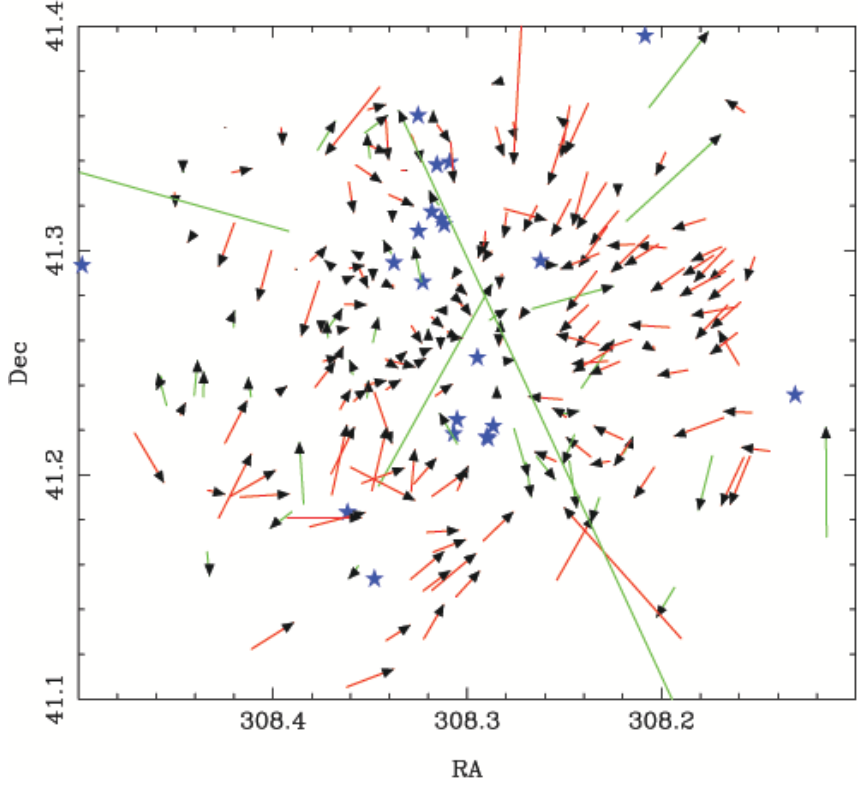}
\caption{PM velocity diagram in the center of Cyg~OB2 (vectors show motion in $10^4$~yrs). Known O-type stars are shown as blue stars. Vectors are color-coded for motion towards (red) or outwards (green) from the central trapezium of O stars (RA = 308.3, DEC = 41.2). These results are preliminary and subject to changes as the analysis is refined.}
\label{vectors}
\end{figure}

\section{Results}

Initial results from 277 stars with PMs and 425 stars with RVs give velocity dispersions of $(\sigma_\alpha, \sigma_\delta, \sigma_r) = (15.6, 12.8, 13.7)$~km/s, equivalent to $\sigma_{3D} = 24.4$~km/s. This implies a virial mass of $M_{dyn} = 9 \times 10^5$~M$_\odot$, significantly larger than the stellar mass, $M_\star = 3 \times 10^4$~M$_\odot$ [21] thereby confirming that the association is gravitationally unbound. Assuming that the stars and the gas were originally in virial equilibrium this implies a star formation efficiency of 3.3\%, a reasonable value.

Figure~1 shows the preliminary distribution of PMs in the center of Cyg~OB2. Immediately apparent is that the majority of stars are moving inwards, not outwards as one would expect for an expanding association. This would appear to argue that infant mortality and tidal stripping are not responsible for disrupting Cyg~OB2, certainly if the association was originally one or two dense clusters. A simple interpretation of the inwards motion is that the association is collapsing under gravity, though this is unlikely given the masses estimated and would require verification from radial velocities. Also of note is the considerable dynamical substructure, with many groups of stars with similar velocities. The substructure suggests the association is not dynamically evolved, despite its age.

\section{Conclusions}

We present 3-dimensional dynamical observations of Cyg~OB2 to elucidate the physical mechanism that led to the association being gravitationally unbound. The overall picture of these observations is a highly substructured, dynamically unmixed association that does not exhibit the position--velocity correlations expected for the theories of infant mortality or tidal stripping. These observations appear to support a picture of hierarchical star formation, in which stars are not born in dense clusters, but in looser associations that retain substructure as they dynamically evolve. Further observations in Cyg~OB2 and other regions are necessary to confirm this.


\end{document}